\documentclass[12pt]{article}
\title{Monte Carlo based Designs for Constrained Domains} 

\renewcommand{\baselinestretch}{1.5}
\usepackage{float}
\usepackage{epsfig}
\usepackage{amsmath}
\usepackage{latexsym}
\usepackage{amssymb}
\usepackage{natbib}
\usepackage{amsthm}
\usepackage{bm}
 \usepackage[english]{babel} 
\usepackage{lmodern}
\usepackage{graphicx}
\usepackage{enumerate}
	\usepackage{fullpage,bbm}
	\usepackage[bookmarks,colorlinks=true,linkcolor=black]{hyperref}
	\usepackage{algpseudocode,algorithm,color}

\usepackage{caption,subcaption}
\theoremstyle{plain}

\theoremstyle{definition}

\newcommand{\bx}{{\bm x}}

\date{}

\author{S. Golchi\thanks{Statistics, University of British Columbia, Kelowna, BC V1V 1V7
		({golchi.shirin@gmail.com}).} 
	\and J.~L. Loeppky \thanks{Statistics, University of British Columbia, Kelowna, BC V1V 1V7}
}

\begin{document}
\maketitle
\renewcommand{\baselinestretch}{1.5}

%{\white

%\begin{centre}
%
%
%\author{Shirin Golchi         \and
%	Jason L. Loeppky%etc.
%}
%
%%\authorrunning{Short form of author list} % if too long for running head
%
%\institute{S. Golchi \at
%	Irving K. Barber School of Arts and Sceinces, Unit 5, University of British Columbia - Okanagan, Kelowna, BC, V1V 1V7\\
%	%              Tel.: +1 778-986-1422\\
%	%              Fax: \\
%	\email{sg3252@columbia.edu}           %  \\
%	%             \emph{Present address:} of F. Author  %  if needed
%	\and
%	J.~L. Loeppky \at
%	Irving K. Barber School of Arts and Sceinces, Unit 5, University of British Columbia - Okanagan, Kelowna, BC, V1V 1V7\\
%}
%
%\end{centre}

%}
%\renewcommand{\baselinestretch}{1.0}

\begin{abstract}
Space filling designs are central to studying complex systems in various areas of science. They are used for obtaining an overall understanding of the behaviour of the response over the input space, model construction and  uncertainty quantification.  In many applications a set of constraints are imposed over the inputs that result in a non-rectangular and sometimes non-convex input space. Many of the existing design construction techniques in the literature rely on a set of candidate points on the target space. Generating a sample on highly constrained regions can be a challenging task. We propose a sampling algorithm based on sequential Monte Carlo that is specifically designed to sample uniformly over constrained regions. In addition, a review of Monte Carlo based design algorithms is provided and the performance of the sampling algorithm as well as selected design methodology is illustrated via examples. 
\noindent
\medskip

\end{abstract}

\renewcommand{\baselinestretch}{1.6}

\section{Introduction}
\label{sec:intro}

The literature on modern design of experiments is rapidly growing due to the demand for efficient design algorithms in many areas of science specifically in cases where computer experiments replace or compliment physical experiments. Computer models are extensively used to simulate complex physical processes. In many cases the computer code is time consuming to run and a surrogate model, such as a Gaussian process (GP) is used to learn about the underlying function using the simulator output \citep{SacWelMit1989,CurMitMor1991,SanWilNot2003}. Since the evaluations of the simulator are the base of inference and prediction, design of computer experiments that refers to appropriately selecting the sample of input values at which the expensive simulator is run is an important first step in computer experiments.

Initial sets of code runs are typically selected using a latin hypercube sample (LHS)\citep{McKBecCon1979}, desirably with some modification such as using a maximin distance \citep{MorMit1995}, correlation based selection criteria \citep{ImaCon1982} or an orthogonal array based LHS \citep{Tan1993,Owe1992}.  Latin hypercube samples can be constructed by assuming the input domain for the computer code is $[0,1]^d$ and assuming a probability distribution function $\mathcal{P}(\bx)\equiv \prod_{i=1}^d \mathcal{P}_i(x_i)$ defined on the hypercube $[0,1]^d$ which is constructed as the product of independent marginal distributions. 

Other design construction strategies for computer experiments are based on optimizing some distance criteria. Minimax and maximin designs \citep{JohMooYlv1990} are two of the most popular distance based designs. Minimax designs minimize the distance between the farthest point in the input space to provide coverage of the input space while maximin designs maximize the minimum distance in the design to spread the design points as far from each other as possible. Both of these criteria become difficult to optimize for a large number of continuous inputs. However, as mentioned above the maximin criterion can be optimized within a class of space-filling designs such as the LHS \citep{MorMit1995}.

While design of computer experiments as well as classical design of experiments have mostly concentrated on rectangular and convex design regions the researchers have recently shifted their attention toward non-rectangular, highly constrained and non-convex regions that appear in various applications \citep{LekJon2014,DraSanDea12,MakJos16,LinShaWin10,PraHarBin16,StiSteHer2003,Tro99,BenHarMul15}. A common first step in a variety of proposed design algorithms both in classical and modern methodology rely on a set of candidate points that provide uniform coverage over the region of interest. Examples are the Fast Flexible space-Filling (FFF) designs of \citep{LekJon2014} who propose hierarchical clustering of a uniform sample over a constrained region and using a summary of the clusters as design points and work of \citep{PraHarBin16} who consider design and analysis of computer experiments with non-convex input spaces by mapping the inputs into a (possibly higher dimensional) space where the Euclidean distances are approximately equivalent to the geodesic distances in the original space.

\begin{figure*}[h!]
	\centering
	\begin{subfigure}[b]{0.48\textwidth}
		\centering
		\includegraphics[width=\textwidth]{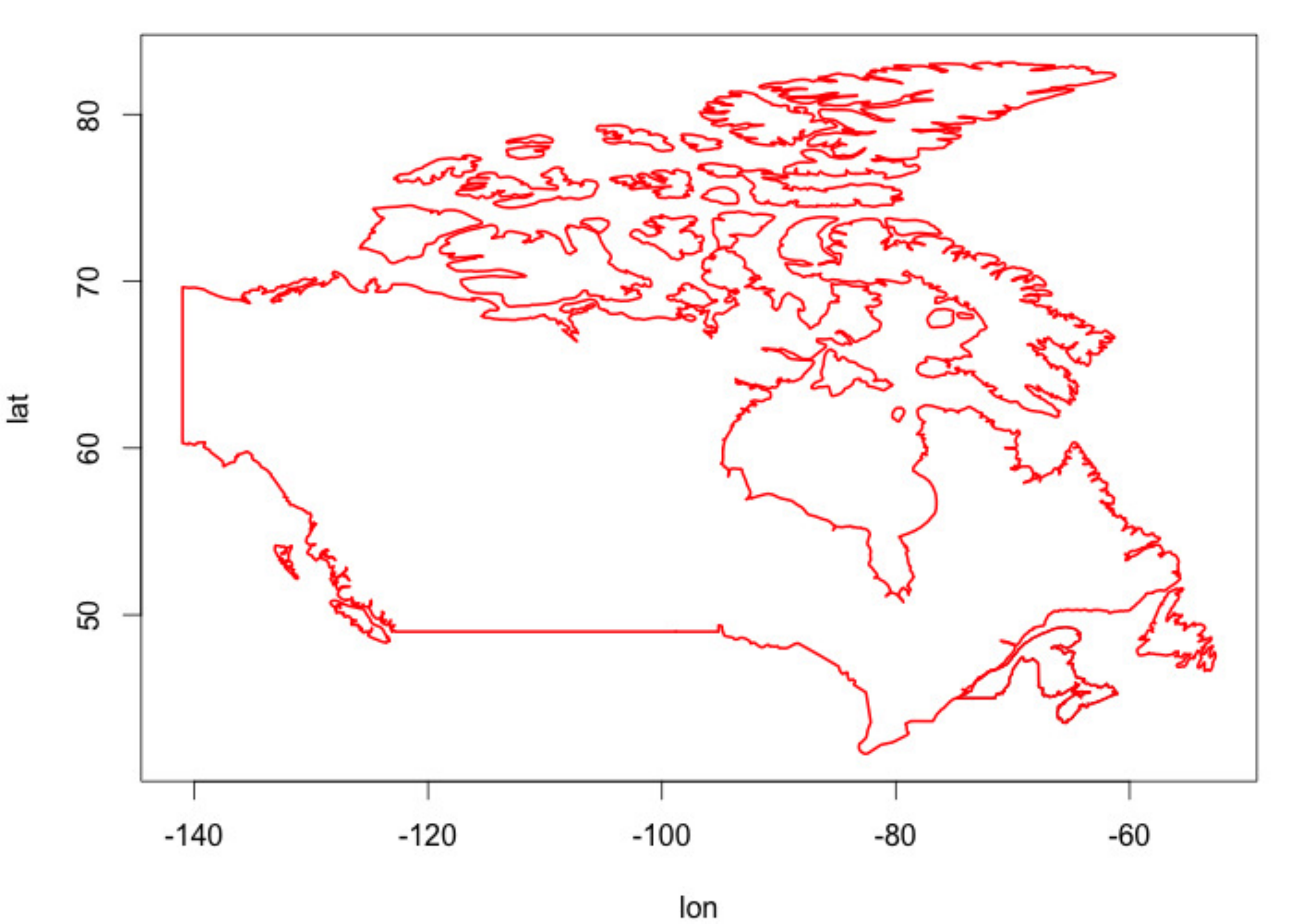}
		\caption{}
		\label{Fig:Canada}
	\end{subfigure}\\
	\begin{subfigure}[b]{0.48\textwidth}
		\centering
		\includegraphics[width=\textwidth]{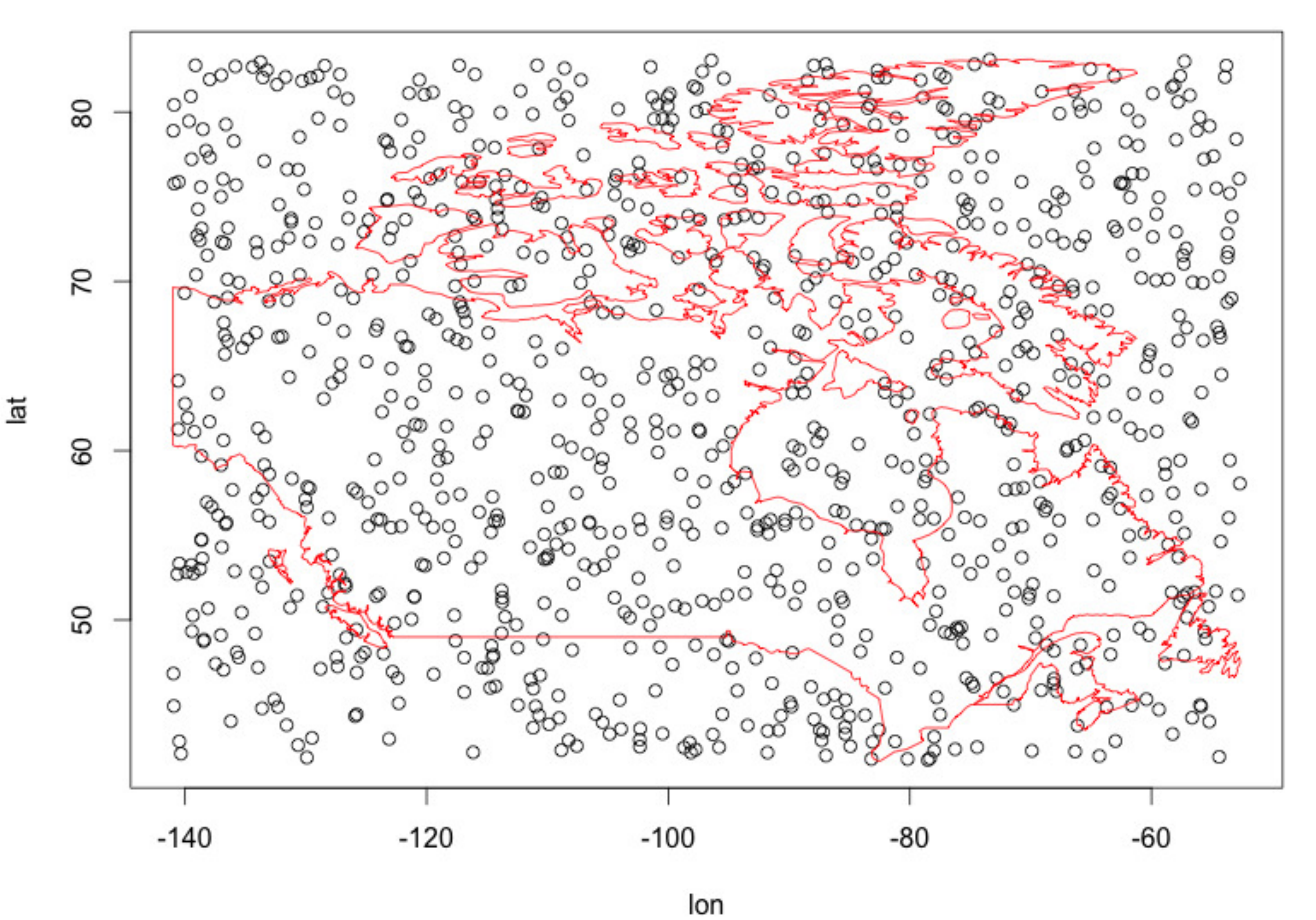}
		\caption{}
		\label{Fig:sample_rectangle}
	\end{subfigure}
	\begin{subfigure}[b]{0.48\textwidth}
		\centering
		\includegraphics[width=\textwidth]{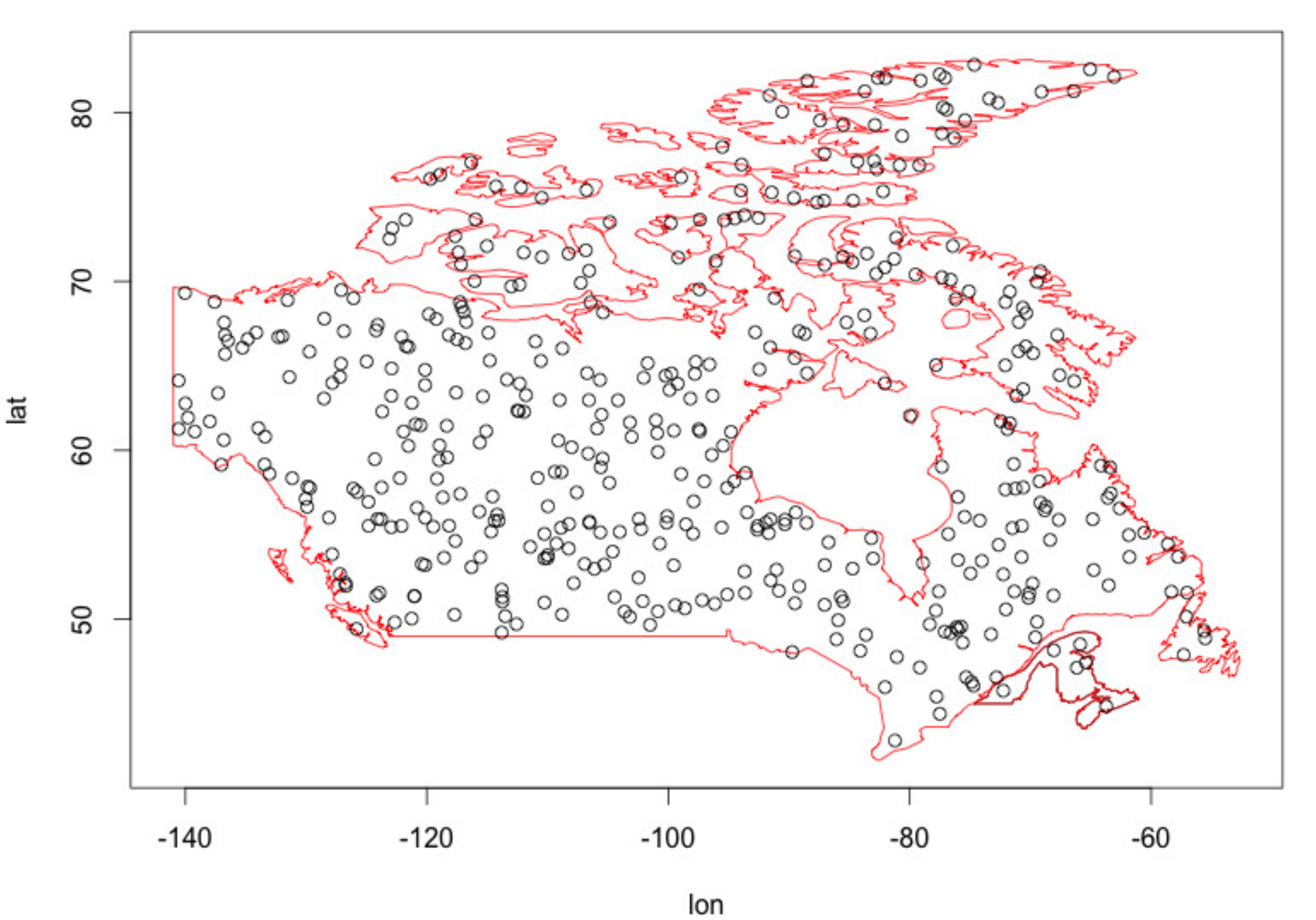}
		\caption{}
		\label{Fig:rejection_sample}
	\end{subfigure}
	\caption{(a) Map of Canada -- an example of a challenging region to generate samples on. (b) A sample of size 1000 randomly generated on a rectangle that contains the map. (c) The remaining sample after discarding the points that fall outside the borders of Canada.}\label{fig:ohCanada}
\end{figure*}

While generating a sample of candidate design points is a trivial step for a rectangular design space it becomes a major challenge for highly constrained regions with non-trivial features. For example consider the map of Canada illustrated in Figure~\ref{Fig:Canada}. The map of Canada comprises one hundred and ten polygons defined by the longitude and latitude of the boundary points. Figure~\ref{Fig:Canada} is in fact a simplified version of the map with only thirty polygons plotted. 

A rejection sampling algorithm may be used to generate a sample from a constrained region. Figure~\ref{Fig:sample_rectangle} shows 1000 points generated on the rectangle that closely contains the map. In Figure~\ref{Fig:rejection_sample} all the points that fall outside the region of interest are discarded. Unfortunately, about half of the initial sample is lost as a result showing the inefficiency of a naive rejection sampler for highly constrained regions. One may resort to Markov chain Monte Carlo sampling in some cases. However, a glance at the map in Figue~\ref{Fig:Canada} reveals the difficulties that an MCMC sampler is likely to face in exploring the space since this problem is equivalent to sampling from a distribution with multiple modes separated by zero probability regions.

In this article we propose a sampling algorithm to generate a uniform sample of candidate points over an arbitrarily constrained region. Our sampling algorithm is based on the sequentially constrained Monte Carlo (SCMC) algorithm proposed by \cite{GolCam2015}. Starting from a uniform sample on the hypercube that contains the target region, SCMC can be formulated to ``move" the sample points inside the constrained region through a combination of importance sampling and MCMC steps without losing samples.

In addition we construct space-filling designs on our example regions using a general design algorithm that relies on sequentially updating a given distance-based or model-based design criterion. The design algorithm layed out and used in this paper is perceived as a general algorithm that covers various existing algorithms in the literature. We provide a brief review of the design criteria that can be adopted into the sequential updating algorithm. The second part of the paper that focuses on design is meant as a review and illustration of a number of methods that are computationally preferred when a set of candidate points is available.

The remainder of the paper is organized as follows. In Section~\ref{sec:SCMC} we explain the sampling algorithm and introduce a general formulation for input constraints. We then illustrate the efficiency and effectiveness of the proposed algorithm through three examples. In Section~\ref{sec:design} we provide a brief review of design algorithms that can be used with a Monte Carlo sample and generate designs on our example regions using some of these design algorithms. Section~\ref{sec:discussion} follows with a discussion and concluding remarks. 

\section{Sampling from constrained input spaces}\label{sec:SCMC}

In this section we explain the adaptation of the sequentially constrained Monte Carlo (SCMC) algorithm to generate a uniform sample across the input space $\mathcal{X}$ that is defined by a set of constraints. These constraints could be inequality constraints that define $\mathcal{X}$ as a $d$-dimensional subspace of $\mathcal{R}^d$, equality constraints that result in a $d$-manifold or a combination of the two types of constraints. The goal is to generate a large uniform sample over $\mathcal{X}$, i.e., to generate a sample from the target distribution
\begin{equation}\label{eqn:target}
\pi^T(\bx)=\frac{\mathcal{P}(\bx)\mathbbm{1}_{\mathcal{X}}(\bx)}{\int_{\mathcal{X}}\mathcal{P}(\bx)d\bx},
\end{equation}
where $\mathcal{P}_{\mathcal{X}}(\bx)$ is the joint probability distribution of the inputs over $\mathcal{X}$ and $\mathbbm{1}_{\mathcal{X}}(\bx)$ is an indicator function that takes a value of 1 if $\bx\in \mathcal{X}$ and is zero otherwise. The normalizing constant of this target distribution, i.e., the integral of $\mathcal{P}_{\mathcal{X}}(\bx)$ over the constrained region, cannot be obtained in most real applications because of the dimensionality and/or complex or unknown form of the constraints. This leaves Monte Carlo schemes that require the distribution to be known only up to a normalizing constant, such as Markov chain Monte Carlo (MCMC) as the only available options for sampling from $\pi^T(\bx)$. However, MCMC is known to be inefficient and difficult to converge when the sampling space is constrained. This is because the sampler has high rejection rates near the boundaries and if the Markov chain is started outside the boundaries of the space, moving away from the zero probability region with a random walk based sampling scheme is extremely challenging. 

We use the approach proposed by \cite{GolCam2015} for sampling from constrained distributions that is based on sequential Monte Carlo. The idea is to relax the constraints fully or to an extent that sampling becomes feasible and use particle filtering schemes to move the samples toward the target distribution while increasing the rigidity of the constraints. A general technique for incorporating explicit soft constraints is proposed in \cite{GolCam2015}. In the following we explain a specialized formulation of this technique for defining probabilistic boundaries for the input space in the constrained design context.

Let us denote the deviation of a given point $\bx$ from the constraints that define the design region $\mathcal{X}$ by $C_{\mathcal{X}}(\bx)$. The deviation function is defined such that its desired value is zero under equality constraints and any negative value under inequality constraints. For example if one is interested in sampling on the unit circle in $\mathcal{R}^2$, i.e., $\mathcal{X}=\{(x_1,x_2): x_1^2+x_2^2=1\}$ the deviation function is defined as 
$$C_{\mathcal{X}}(x_1,x_2)=|x_1^2+x_2^2-1|,$$
while if we are interested in sampling inside the unit circle in $\mathcal{R}^2$, i.e., $\mathcal{X}=\{(x_1,x_2): x_1^2+x_2^2<1\}$ the deviation function is defined as
$$C_{\mathcal{X}}(x_1,x_2)=x_1^2+x_2^2-1.$$
The following probit function is a probabilistic version of the on/off constraint indicator
\begin{equation}\label{eqn:profit}
\Phi(-\tau C_{\mathcal{X}}(\bx)),
\end{equation}
where $\Phi$ is the normal cumulative distribution function and the parameter $\tau$ controls the slope of the probit function or in other words the strictness of the constraint. The above function converges to the strict constraint indicator in the limit.
\begin{equation}\label{eqn:profit}
\lim_{\tau\rightarrow \infty}\Phi(-\tau C_{\mathcal{X}}(\bx))=\mathbbm{1}_{\mathcal{X}}(\bx).
\end{equation}
This parametrization of the constraints allows us to define a sequence of densities that impose the constraints more strictly at each step moving toward the target space,
\begin{equation}\label{eqn:seq}
\{\pi^t(\bx)\}_{t=0}^{T},
\end{equation}
\begin{equation}
\pi^t(\bx)\propto \mathcal{P}(\bx)\Phi(-\tau_t C_{\mathcal{X}}(\bx)),
\end{equation} 
\begin{equation}
0=\tau_0<\tau_1<\ldots<\tau_T\rightarrow \infty,
\end{equation}
If $\mathcal{X}$ is defined by a set of constraints (i.e., $C_{\mathcal{X}}(\bx)=\{C_k(\bx)\}_{k=1}^{K}$) the intermediate densities are defined as,
\begin{equation}
\pi^t(\bx)\propto \mathcal{P}(\bx)\prod_{k=1}^K\Phi(-\tau_t C_k(\bx)).
\end{equation}
Note that equality constraints in a continuous parameter space are satisfied with probability zero. However, the deviation from the constraint can be made arbitrarily small by the choice of the final value of the constraint parameter, $\tau_T$. 

The SCMC sampler is used to draw a sample from $\mathcal{P}(\bx)$ and filter this initial sample sequentially toward $\mathcal{X}$. Algorithm~\ref{alg:SCMC} outlines the SCMC sampler for the case that the inputs are assumed to be distributed uniformly and independently. The initial sample in this case is a uniform sample over the d-dimensional hypercube $\mathcal{Q}^d$ that contains $\mathcal{X}$. Note that being able to define $\mathcal{Q}^d$ as close as possible to the target space results in more efficiency of the SCMC sampler. 

An important decision to be made for any SMC sampler is the distance between two consecutive densities in the sequence as well as the length of the sequence. These two factors highly affect the efficiency of the sampler. In our framework this decision reduces to choosing an effective sequence of constraint parameters $\tau_t$. This is achieved adaptively in step 3(a) of Algorithm~\ref{alg:SCMC} that is inspired by \cite{JasSteDou11} who proposed an adaptive approach for a specific family of SMC algorithms. The idea is to determine the next density in the sequence such that the effective sample size (ESS) does not fall below a given value (for example half of the sample size) in transition from one density to the other. This is done by numerically solving the following equation for $\tau_t$,
\begin{equation}
\label{eqn:adapt_ESS}
\text{ESS}=\frac{\left(\sum_{n=1}^N w^{t}_n(\tau_t)\right)^2}{\sum_{n=1}^N\left(w^{t}_n(\tau_t)\right)^2},
\end{equation}
where
\begin{equation}
\label{eqn:adapt_weight}
w^{t}_n(\tau_t)=\frac{\Phi(-\tau_t C_{\mathcal{X}}(\bx_n^{t-1}))}{\Phi(-\tau_{t-1} C_{\mathcal{X}}(\bx_n^{t-1}))}.
\end{equation}
The length of the sequence is also determined by this adaptive approach: a target value for the constraint parameter is chosen (e.g. $10^6$) and the algorithm is run until the predetermined value of the ESS is achieved by this target value.

Another key step of the SCMC algorithm is the sampling step (step 3(h)) that prevents particle degeneracy. If the sampling step is skipped or is not done effectively a few probable points are repeatedly copied in the resampling step and in the end one might be left with a small proportion of distinct points in the input space. The sampling step comprises one (or a few) MCMC transition step(s) that move the samples slightly under the current posterior at time $t$, i.e., a proposal followed by a Metropolis-Hastings accept/reject step for each sample point. The efficiency of the algorithm can be improved by using the covariance structure of the current density in the proposal distribution if such information is available. However, as a general guideline we suggest Gibbs type transitions (i.e., one dimension at a time) with normal proposal distributions whose variances are chosen adaptively by monitoring the acceptance rate from the previous time step. The notation $\bx^{t}_n\sim K^{t}$ is used to show the Gibbs/Metropolis-Hastings step for a sample point $\bx_n$ where $K^t$ is a transition kernel for $\pi^t$.

\begin{algorithm}[t]
	\caption{SCMC sampling from the space $\mathcal{X}$}\label{alg:SCMC}
	\begin{algorithmic}[1]
	\renewcommand{\algorithmicrequire}{\textbf{Inputs:}}
	\renewcommand{\algorithmicensure}{\textbf{Return:}}
	\Require Hypercube $\mathcal{Q}^d\supset \mathcal{X}$, constraint $C_{\mathcal{X}}$, sample size $N$, Target constraint parameter value $\tau_T$.\\
	
	$t\gets 0$\\
	
	Generate a uniform sample $S$ of size $N$ on $\mathcal{Q}^d$;\\
	
	Initiate the weights $W_n^0\gets \frac{1}{N}$ for $n=1,\ldots,N$;
	
	\While  {$\tau_t\leq \tau_T$} 
	\begin{enumerate}[(a)]
	\item $t\gets t+1$
	
	\item Numerically solve~(\ref{eqn:adapt_ESS}) to obtain $\tau_t$;
	
	\item $W^t_n\gets W_{t-1}^n w^t_n$ where $w^t_n=\frac{\Phi(-\tau_t C_{\mathcal{X}}(\bx_n^{t-1}))}{\Phi(-\tau_{t-1} C_{\mathcal{X}}(\bx_n^{t-1}))}$, $n=1,\ldots,N$;
	
    \item Normalize $W^{t}_{1:N}$, i.e., $W^t_n\gets \frac{W^t_n}{\sum_{n=1}^{N}W^t_n}$;
	
	\item Resample $\bx_{1:N}^{t-1}$ with weights $W^{t}_{1:N}$; 
	    
	\item $W^{t}_{1:N}\gets \frac{1}{N}$;
	
	\item Sample $\bx^{t}_{1:N}\sim K^{t}$ (refer to the text for $K^t$);
	
	\end{enumerate}
	\EndWhile
	
	\Ensure Uniform sample $\bx^T_{1:N}$ from $\mathcal{X}$.
	\end{algorithmic}
	\end{algorithm}

In the following we demonstrate the performance of the sampling algorithm via three example s of constrained regions. Our first example is a non-convex but simple subset of $\mathcal{R}^2$ that is used to illustrate the implementation of SCMC as well as the evolution of the sample through the steps of the sampling algorithm. The other examples are chosen to demonstrate the performance of the algorithm in more challenging problems with non-trivial constraints.
	
\begin{figure*}[h!]
	\centering
	\begin{subfigure}[b]{0.45\textwidth}
		\centering
		\includegraphics[width=\textwidth]{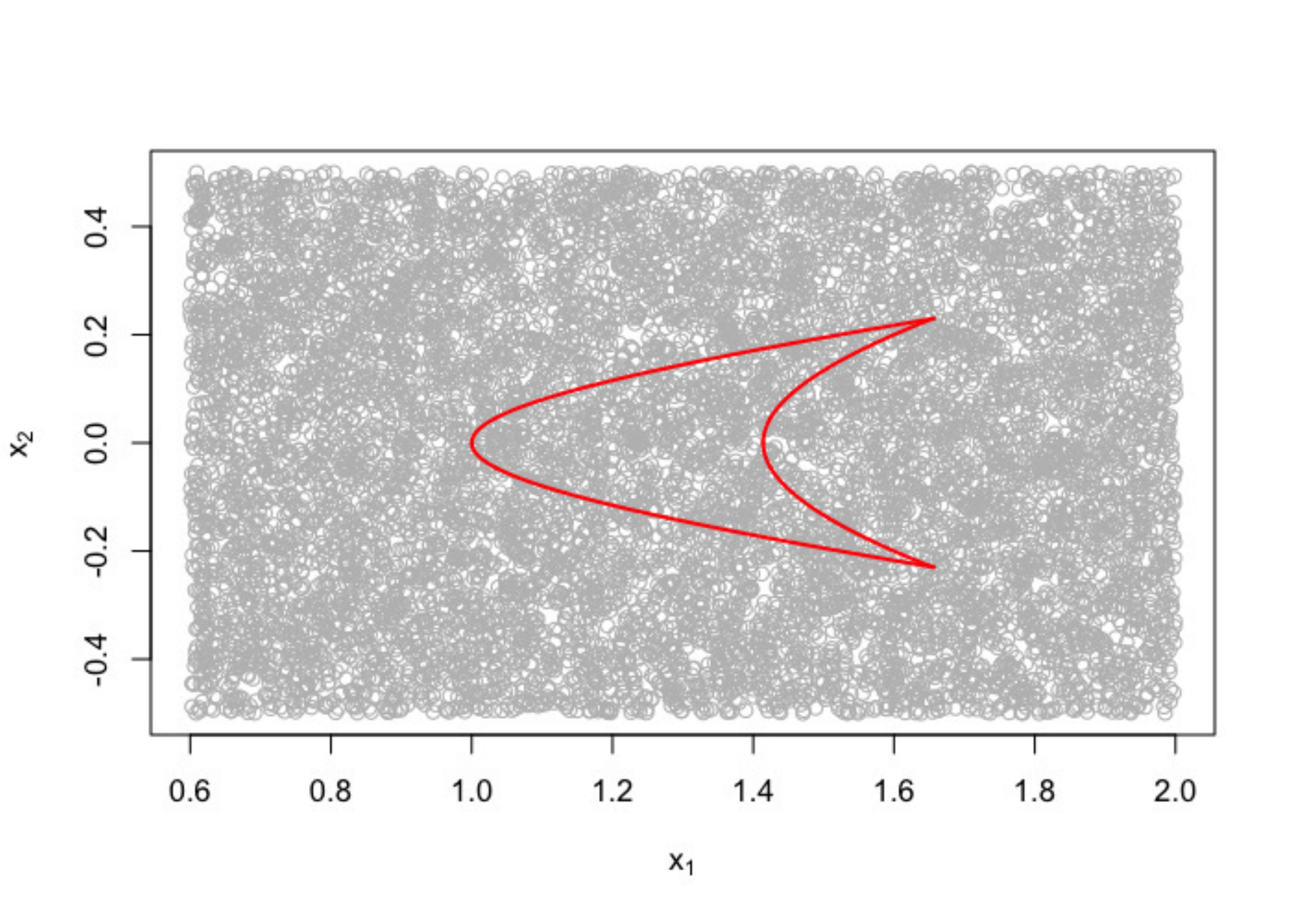}
		\caption{$t=0$, $\tau_t=0$}
		\label{Fig:cres1}
	\end{subfigure}
	\begin{subfigure}[b]{0.45\textwidth}
		\centering
		\includegraphics[width=\textwidth]{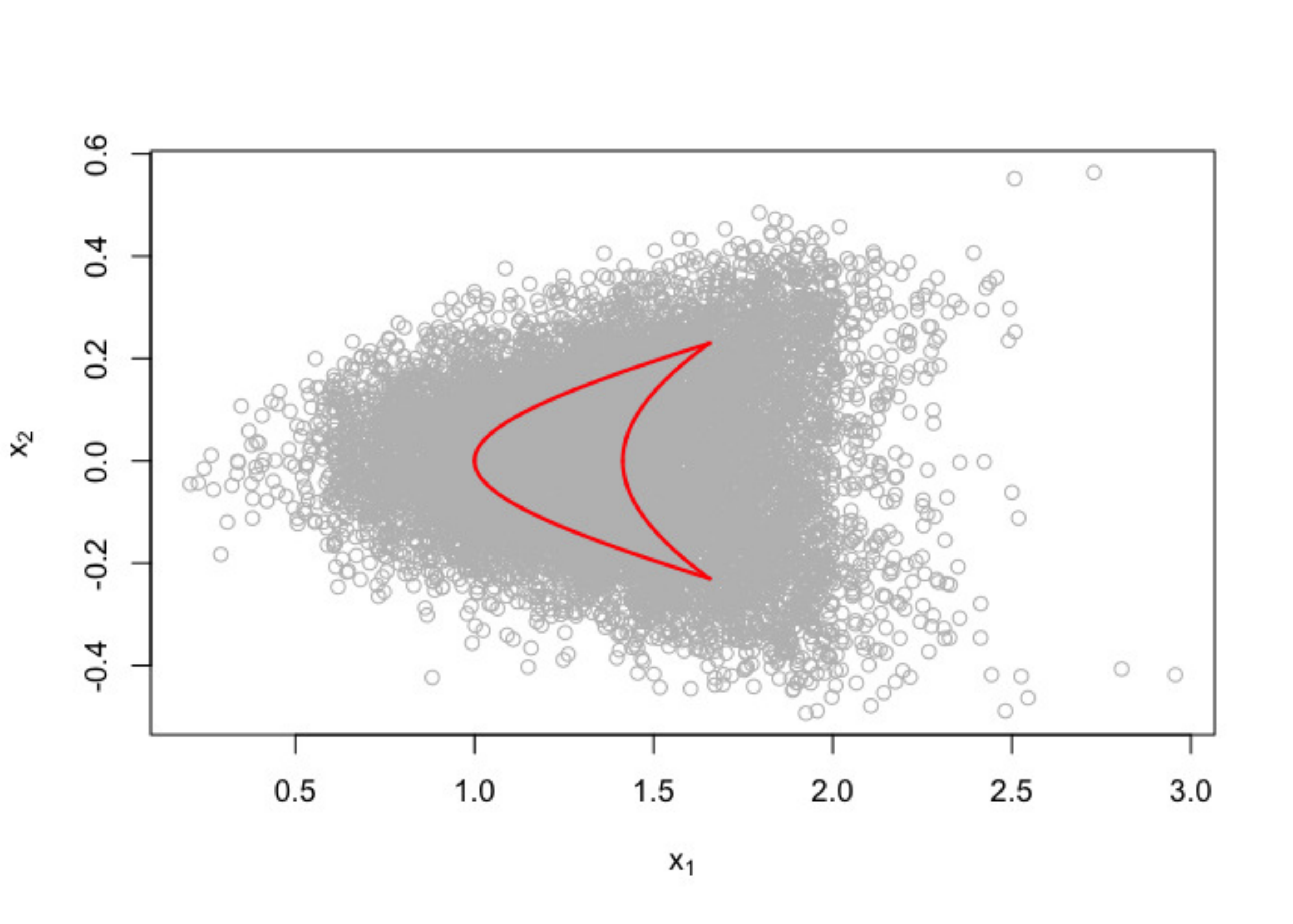}
		\caption{$t=1$, $\tau_t=2.33$}
		\label{Fig:cres2}
	\end{subfigure}
	\begin{subfigure}[b]{0.45\textwidth}
		\centering
		\includegraphics[width=\textwidth]{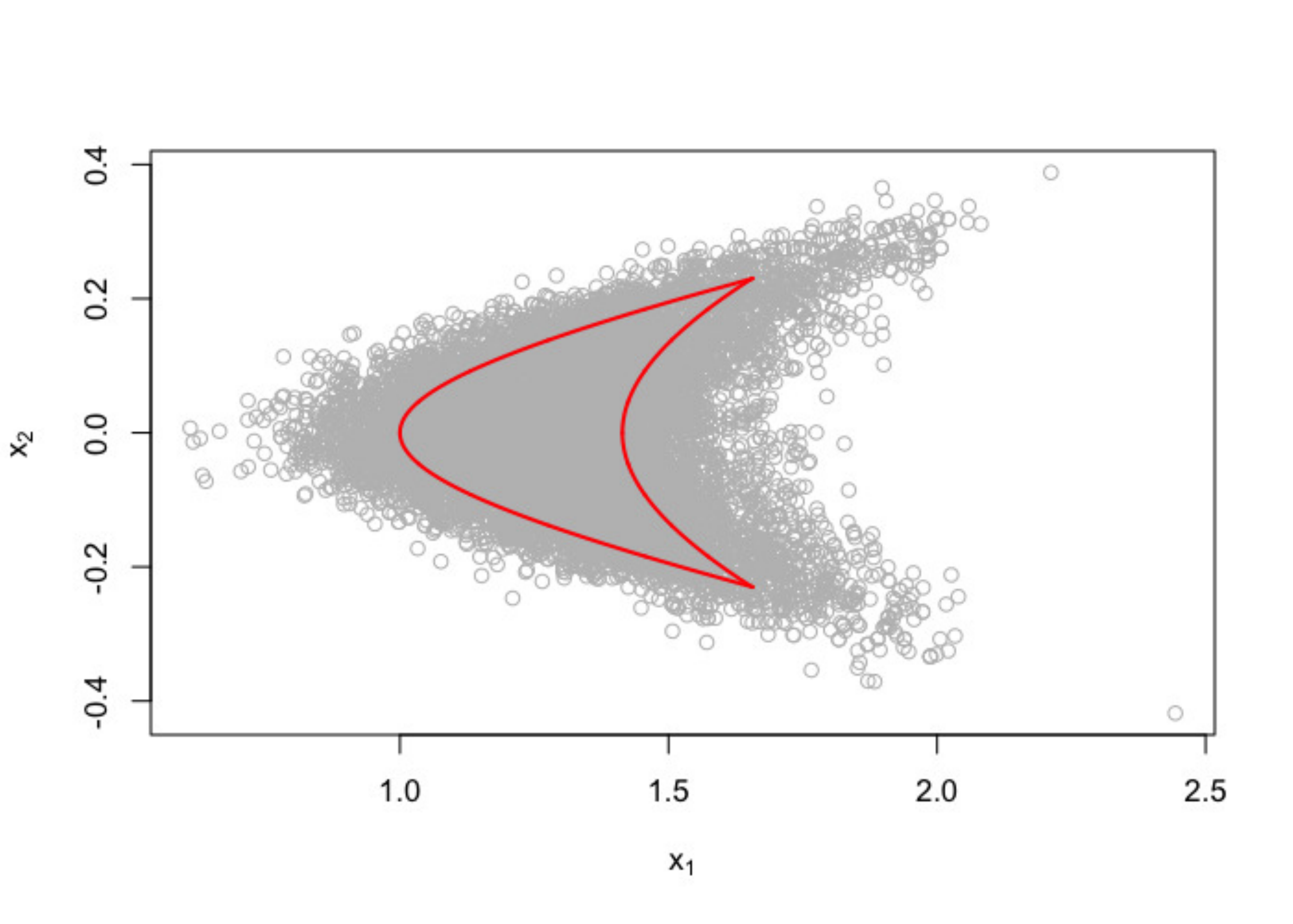}
		\caption{$t=2$, $\tau_t=6.95$}
		\label{Fig:cres3}
	\end{subfigure}
	\begin{subfigure}[b]{0.45\textwidth}
		\centering
		\includegraphics[width=\textwidth]{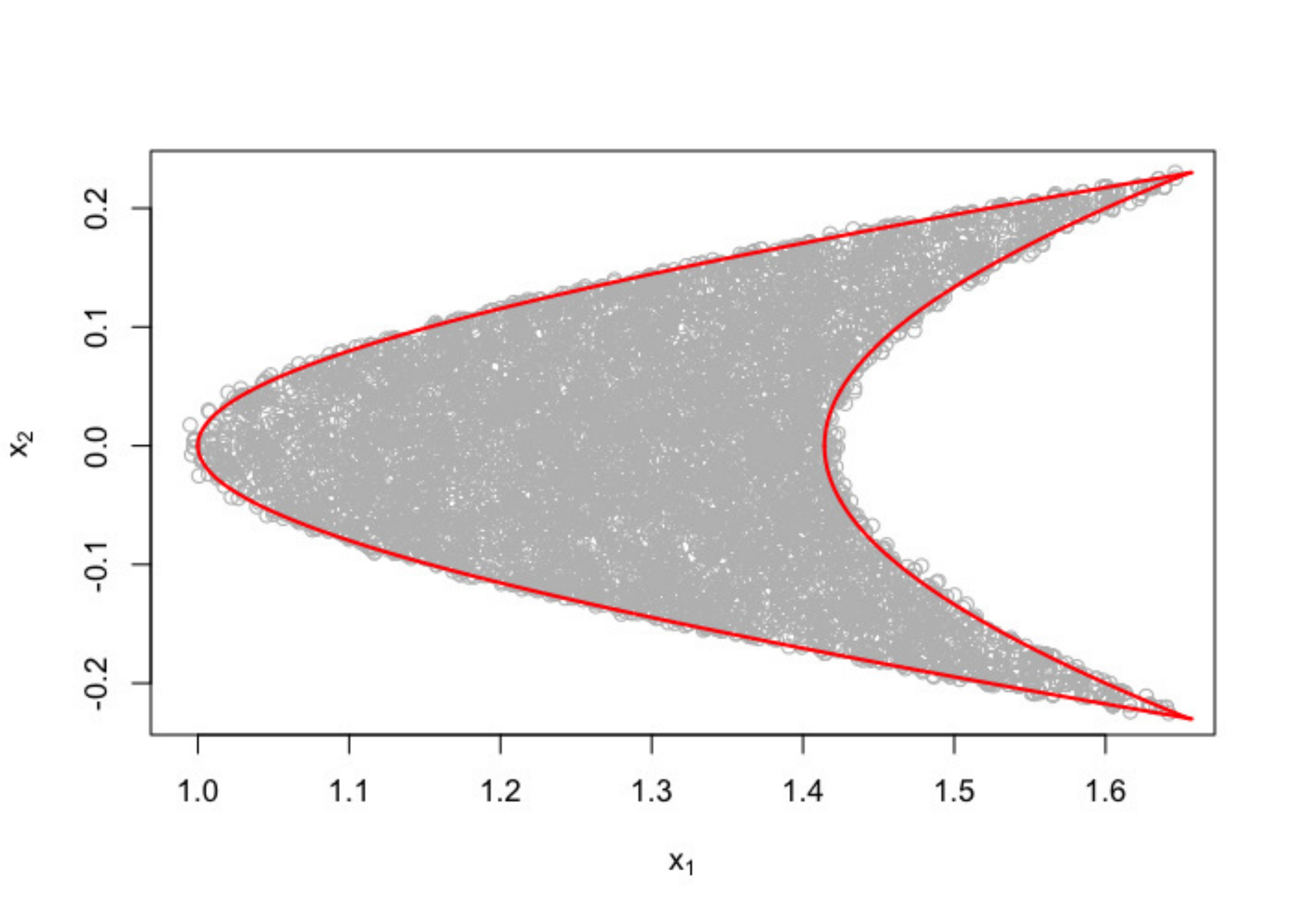}
		\caption{$t=3$, $\tau_t=1.25\times 10^2$}
		\label{Fig:cres4}
	\end{subfigure}
	\begin{subfigure}[b]{0.45\textwidth}
		\centering
		\includegraphics[width=\textwidth]{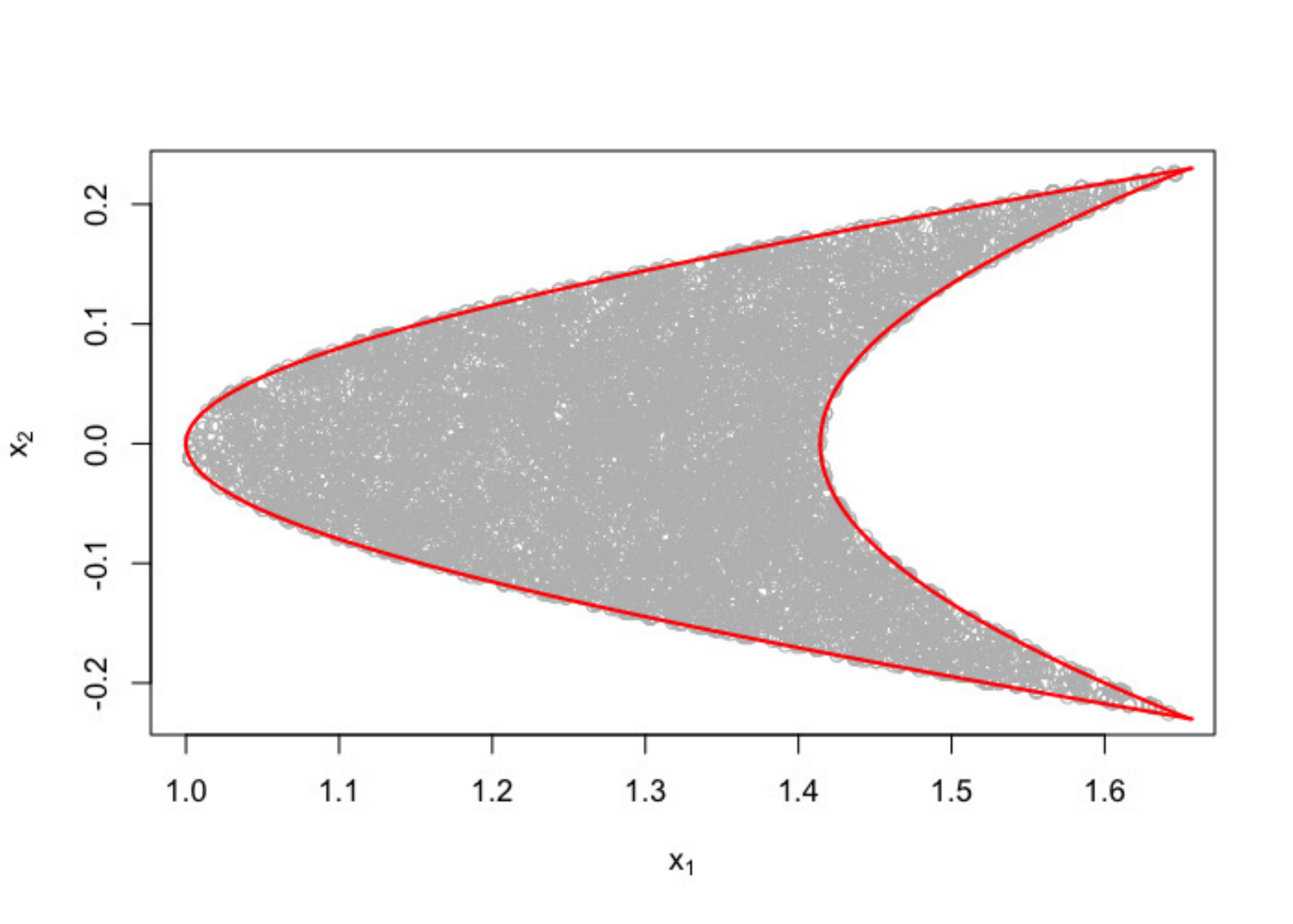}
		\caption{$t=4$, $\tau_t=1\times 10^6$}
		\label{Fig:cres5}
	\end{subfigure}
	\caption{The evolution of the sample of points from the unconstrained hypercube to the constrained region.}\label{Fig:crescent_evolve}
\end{figure*}

{\bf Example 1.} Consider a non-convex subset of $\mathcal{R}^2$ defined by the following non-linear inequality constraints, 
\begin{equation}\label{eqn:crescent}
\mathcal{X}=\left\{(x_1,x_2)\in \mathcal{R}^2, \sqrt{33x_2^2+1}<x_1<\sqrt{14x_2^2+2}\right\}.
\end{equation}
The deviation vector is given by 
\begin{equation}
	C_\mathcal{X}(x_1,x_2)= \begin{pmatrix}
	 	x_1 - \sqrt{14x_2^2+2} \\
	 	\sqrt{33x_2^2+1} - x_1 
	 \end{pmatrix}
\end{equation}
and the target distribution is defined as,
\begin{equation}
\pi_T \propto \Phi\left(-\tau\left(x_1 - \sqrt{14x_2^2+2}\right)\right)\Phi\left(-\tau\left(\sqrt{33x_2^2+1} - x_1 \right)\right)
\end{equation}
Algorithm~\ref{alg:SCMC} is used to generate a uniform sample of size 10,000 over $\mathcal{X}$. With a final constraint parameter value of $\tau_T=10^6$ the algorithm terminates at $t=4$. Figure~\ref{Fig:crescent_evolve} shows the evolution of the sample in five time steps.

As can be seen in Figure~\ref{Fig:cres1}, we are overly conservative in this example by starting the sampling on a rectangle that is much larger than the constrained region. However, despite the conservative first step the sampler is able to converge to the target space very quickly.

\begin{figure*}[t]
	\centering
	\begin{subfigure}[b]{0.45\textwidth}
		\centering
		\includegraphics[width=\textwidth]{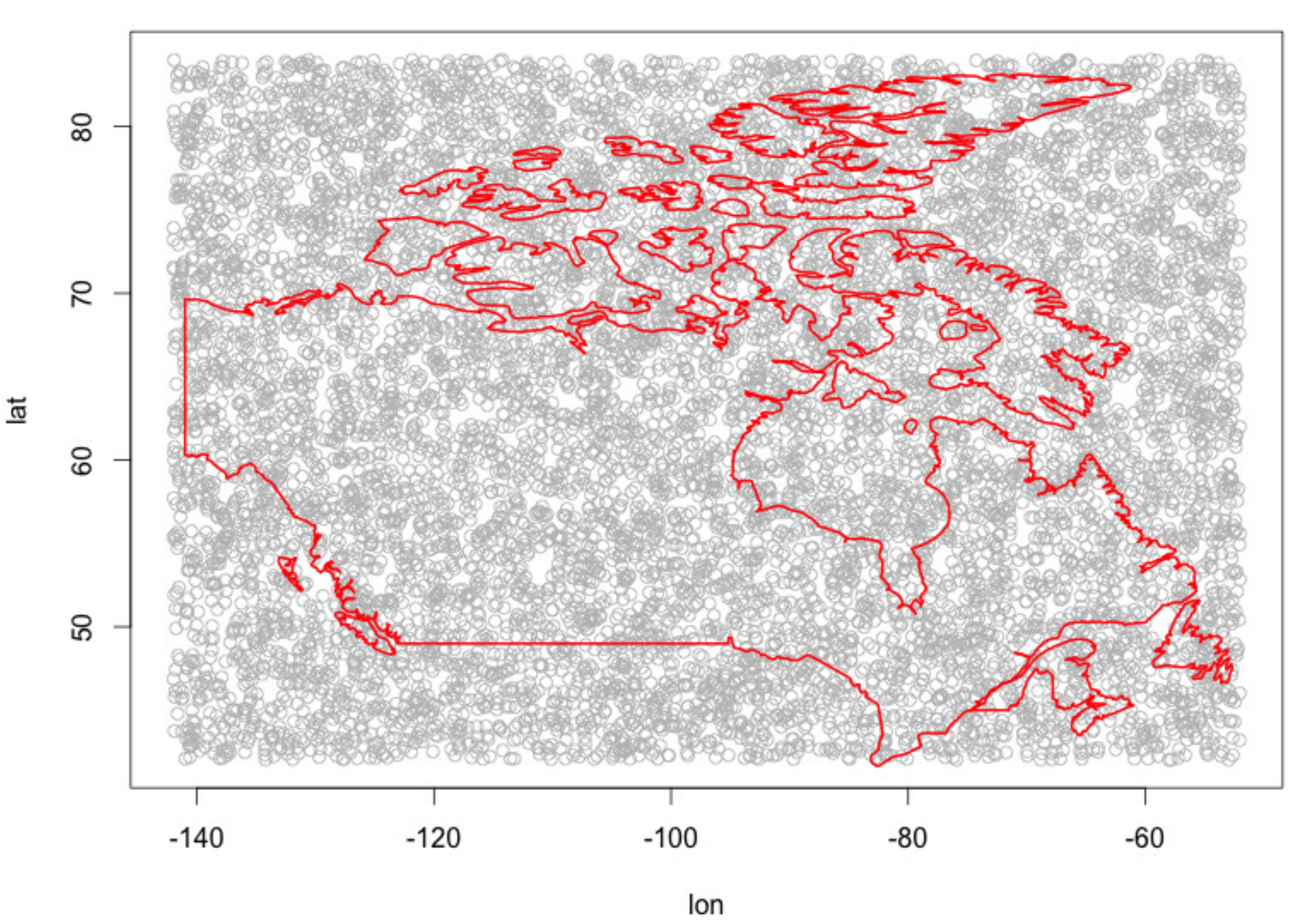}
		\caption{$\tau_t=0$}
		\label{Fig:canada1}
	\end{subfigure}
	\begin{subfigure}[b]{0.45\textwidth}
		\centering
		\includegraphics[width=\textwidth]{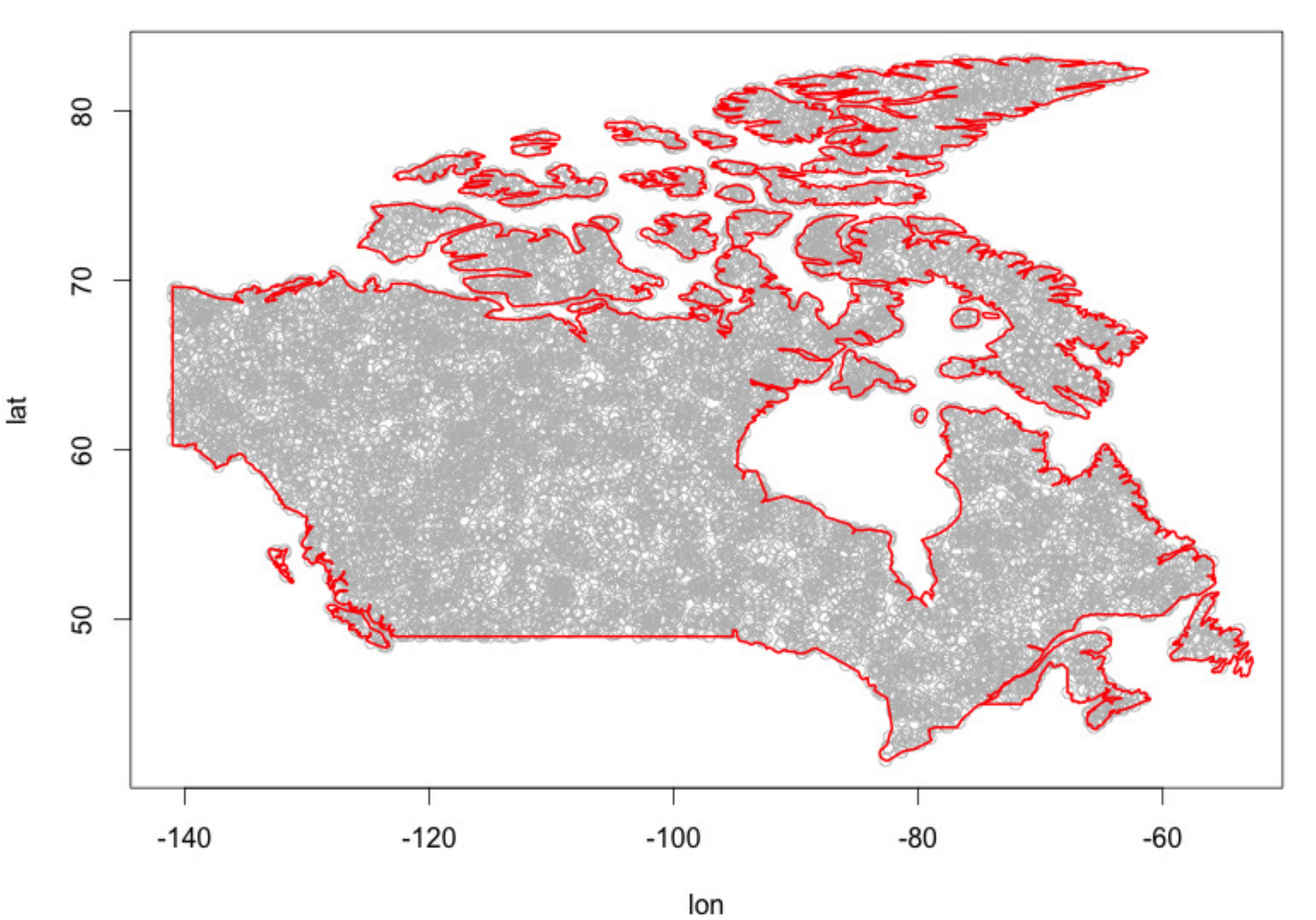}
		\caption{$\tau_t=1\times 10^6$}
		\label{Fig:canadaT}
	\end{subfigure}
	\caption{The evolution of the sample of points from the unconstrained hypercube to the constrained region.}\label{fig:canada_sample}
\end{figure*}

{\bf Example 2.} We now revisit the map of Canada that was used as a motivating example in Section~\ref{sec:intro}. To use the SCMC sampler to generate a sample over the map we need to define a deviation function. The most trivial deviation function is the distance to the set that defines Canada for any arbitrary point $\mathbf{x}$ on the planet specified by its longitude and latitude. Denoting the longitude by $x_1$ and the latitude by $x_2$ the deviation function is given by
\begin{equation*}
C_{\mathcal{X}}(\mathbf{x})=\begin{cases}0 & \text{if } \mathbf{x}\in \mathcal{X} \\
\min_{\mathbf{u}\in \mathcal{X}}\delta(\mathbf{x},\mathbf{u}) & \text{if } \mathbf{x}\notin \mathcal{X}\end{cases}
\end{equation*}
 where $\delta$ is the Euclidean distance. As mentioned before the map of Canada comprises one hundred and ten polygons which results in higher cost in computation of the deviation function. Therefore, we use a simplified version of the map with only thirty polygons. Figure~\ref{fig:canada_sample} shows 100,000 samples at the initial and final steps of the SCMC sampler respectively.

\begin{figure*}[t]
	\centering
	\begin{subfigure}[b]{0.45\textwidth}
		\centering
		\includegraphics[width=\textwidth]{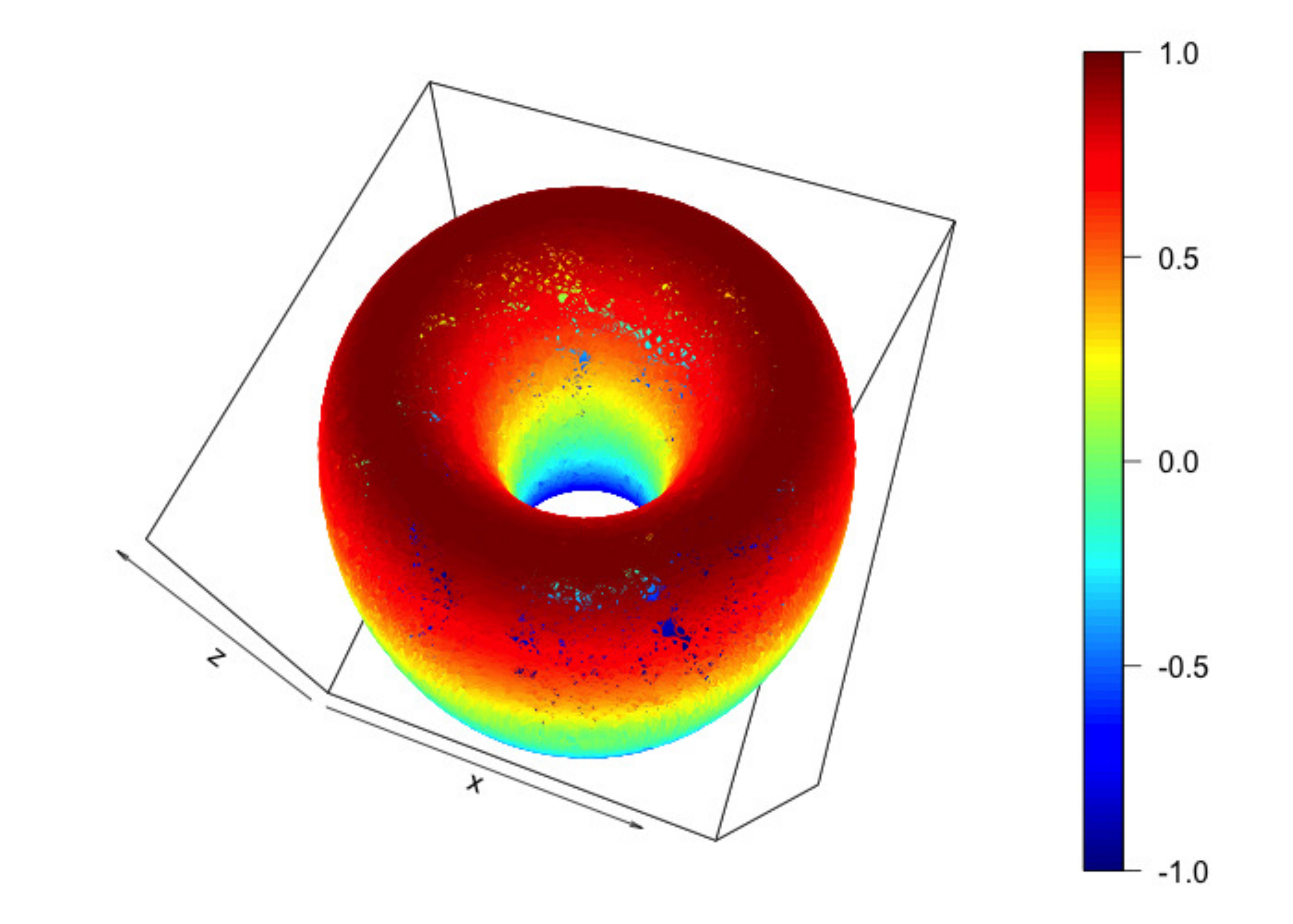}
		\caption{}
		\label{Fig:canada1}
	\end{subfigure}
	\begin{subfigure}[b]{0.45\textwidth}
		\centering
		\includegraphics[width=\textwidth]{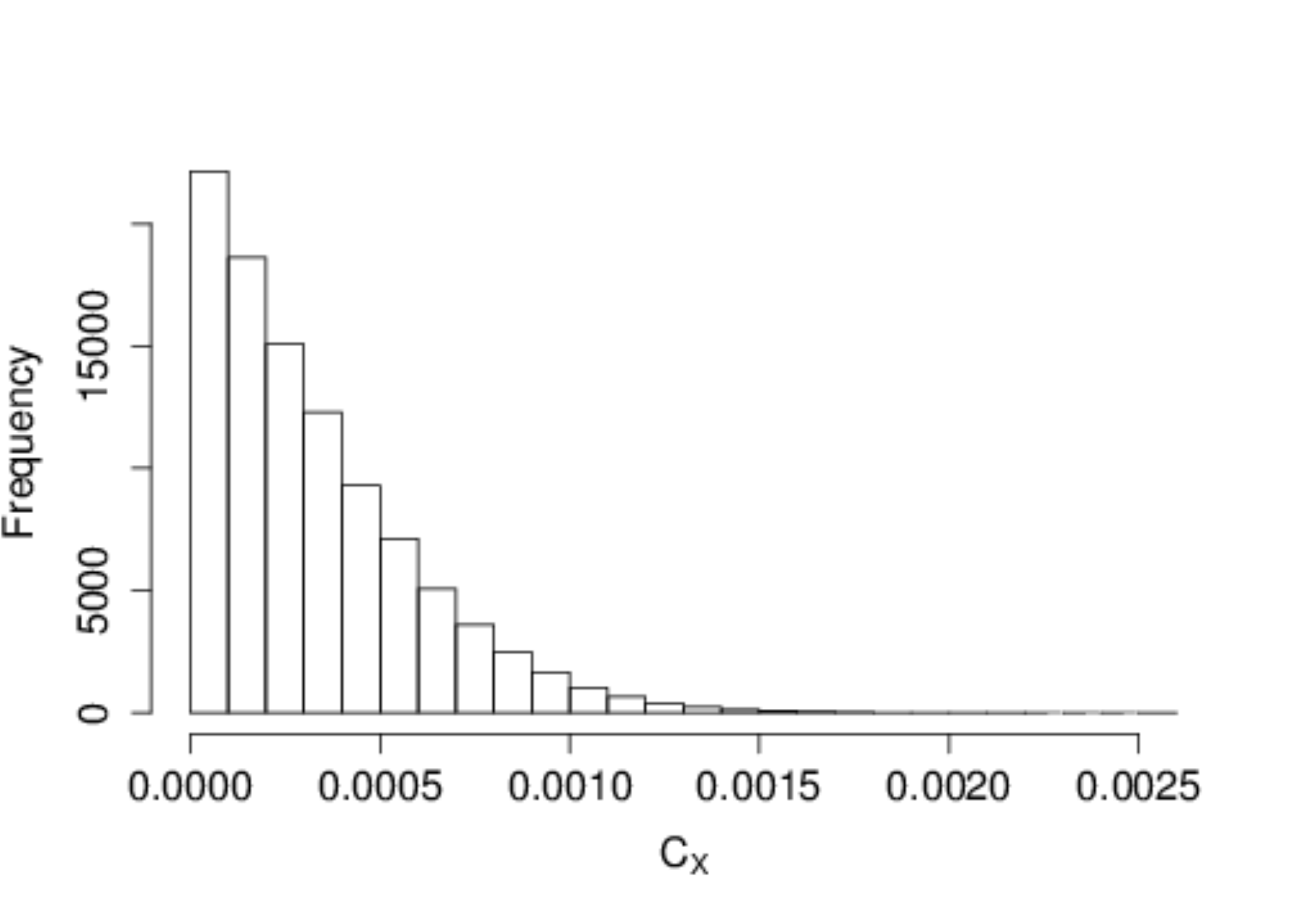}
		\caption{}
		\label{Fig:canadaT}
	\end{subfigure}
	\caption{(a) Sample of points generated on (close to) the torus; (b) histogram of the deviation of the sample points from the torus.}\label{fig:torus}
\end{figure*}

{\bf Example 3.} As the last example we consider a manifold in $\mathcal{R}^3$. The target space is a torus given by 
\begin{equation*}\label{eqn:torus}
\left(2-\sqrt{x_1^2+x_2^2}\right)^2+x_3^2=1.
\end{equation*}
Generating samples on manifolds using Monte Carlo methods in the embedding space is an unsolved problem \footnote{\href{https://xianblog.wordpress.com/2014/03/24/mcmc-on-zero-measure-sets/}{https://xianblog.wordpress.com/2014/03/24/mcmc-on-zero-measure-sets/}}.  However, as noted by \cite{GolCam2015} SCMC generates samples arbitrarily close to the manifold. Running the SCMC algorithm with the constraint function
\begin{equation*}\label{eqn:torus}
C_{\mathcal{X}}(x_1,x_2,x_3)=|\left(2-\sqrt{x_1^2+x_2^2}\right)^2+x_3^2-1|
\end{equation*}
and a 100,000 samples results in a sample of points with a maximum deviation of $.0026$. Figure~\ref{fig:torus} shows the sample together with the histogram of the corresponding deviations.

\section{Monte-Carlo-based designs}\label{sec:design}

In this section we  discuss a number of methods to efficiently generate a space filling design on an input space $\mathcal{X}$ using a large uniform sample $S=\{\bx_1,\ldots,\bx_N\}$ over $\mathcal{X}$. We consider a general framework for design construction that relies on adding points one (or a batch) at a time by sequentially optimizing a distance or model based design criterion. We generate designs on the example regions introduced in the previous section using the sequential optimization with the maximin criterion as well as the FFF design algorithm of \cite{LekJon2014}.

\subsection{Conditionally Optimal Designs}
Let us denote the sequential space filling design we wish to construct on $\mathcal{X}$ by $s=\{\bx^1,\ldots,\bx^P\}$ where $\bx^p$ is the design point selected at step $p=1,\ldots,P$. The design $s$ is generated by sequentially selecting points from $S$ that maximize  a conditional criterion defined in compliance with an overall distance-based or model-based measure of performance for the design,
\begin{equation}\label{eqn:criterion}
\Psi(\cdot,s^{p-1}),
\end{equation}
where $s^{p-1}$ is the design up to selection of the $p^{\text{th}}$ point.
 
Various design algorithms in the literature fall under this general framework. For example, \cite{KenSto69}, proposed constructing maximin designs by sequentially adding points that maximizing the minimum distance in the design. A design $s$ is maximin if it maximizes $\min_{\bx_i,\bx_j\in s}\delta(\bx_i,\bx_j)$ \citep{JohMooYlv1990}. The conditional maximin criterion is defined as,
\begin{equation}\label{eqn:MmmM}
\Psi_\delta(\cdot,s^p)=\min_{\bx_j\in s^p}\delta(\cdot,\bx_j).
\end{equation}
A design obtained by iteratively maximizing~(\ref{eqn:MmmM}) is, by definition, a conditional maximin design (cMm). At each step $p$, the design $s^p$ is the maximum distance design among those designs that contain $s^{p-1}$ resulting in an increasing sequence of cMm designs,
$$s^1\subset s^2\subset \ldots \subset s^P.$$

The cMm designs can be obtained with a computationally efficient algorithm outlined in Algorithm~\ref{alg:design}. The efficiency of this algorithm is a result of the nature of the maximin criterion: at each step $p$ the minimum distance in the $p$-point maximin design is the distance of the most recently added point and the closest design point. This is because the $p$-point design is obtained by adding a point to a $(p-1)$-point maximin design and the minimum distance can only decrease at each step. This property saves the computation of the minimum distance of all the $N-p$ candidate designs at step $p$. The maximin design is obtained by adding the point in $S$ that has the maximum distance to the design. The distance of a point to the design at each step is also obtained by computing the distance with the most recently added point in the design since the distances with the rest of the design points are already computed in the previous iterations. 

Other distance based design criteria can also be optimized sequentially in a computationally efficient way.  In fact, any design criterion that depends on pairwise distances would be more cost efficient to compute and optimize in the conditional optimization framework. Consider for example the  non-collapsing space filling designs proposed by \cite{DraSanDea12} that is based on average reciprocal distance (ARD) that focuses on distance performance in lower dimensional projections as originally discussed in  \cite{WelBucSac1996}. For a design $s^{p-1}$ with $p-1$ points the ARD design criterion is given by 
\begin{equation}\label{eqn:ARD}
\Psi(s^{p-1})=\left\{ \frac{1}{\sum_{q\in{1,\ldots,p-1}}{{p-1}\choose q}}\sum_{q\in{1,\ldots,p-1}}\sum_{r=1}^{{p-1}\choose q}\sum_{\bx_i,\bx_j\in s^{p-1}}\frac{q^{\frac{k}{2}}}{\delta^k_{qr}(\bx_i,\bx_j)}\right\}^{-\frac{1}{k}}.
\end{equation}
where $\delta^k_{qr}$ is the $k^{\text{th}}$ order Euclidean distance in the $r^{\text{th}}$ projection of the $q^{\text{th}}$ dimension. This criterion becomes very expensive to compute as the dimensionality of the input space increases. Suppose that $s^p=\{\bx^*\}\cup s^{p-1}$ the ARD design criterion for $s^p$ is given by
\begin{equation}\label{eqn:ARD}
\Psi(s^p)=\left\{\Psi(s^{p-1})^{-k} + \frac{1}{\sum_{q\in{1,\ldots,p}}{p\choose q}}\sum_{q\in{1,\ldots,p}}\sum_{r=1}^{p\choose q}\sum_{\bx_j\in s^{p-1}}\frac{q^{\frac{k}{2}}}{\delta^k_{qr}(\bx^*,\bx_j)}\right\}^{-\frac{1}{k}}.
\end{equation}
Therefore to conditionally maximize the ARD criterion we need to minimize 
\begin{equation}\label{eqn:ARD}
\frac{1}{\sum_{q\in{1,\ldots,p}}{p\choose q}}\sum_{q\in{1,\ldots,p}}\sum_{r=1}^{p\choose q}\sum_{\bx_j\in s^{p-1}}\frac{q^{\frac{k}{2}}}{\delta^k_{qr}(\bx^*,\bx_j)}
\end{equation}
 which requires calculating the distances between the candidate points and the design points rather than all the pairwise distances as a result of adding a new point. Note that at each step the distances calculated between the candidate points and the design points up to the previous step can be recycled. So one only needs to calculate the distance between the candidate points and the most recently added design point which reduces the computation cost significantly. 

Another recently introduced design criterion is the maximum projection (MaxPro) criterion by \cite{JosGul15} that is the average reciprocal product of squared one-dimensional distances. The sequential equivalent of MaxPro criterion is given by,
\begin{equation}\label{eqn:ARD}
\Psi(\cdot \mid S^{p-1})=\left \{\frac{1}{p-1}\sum_{\mathbf{x}_j\in s^{p-1}}\frac{1}{\prod_{d=1}^{D}\delta^2(\cdot , x_{jd})}\right\}^{\frac{1}{D}}.
\end{equation}
which is the average reciprocal product of 1-d distances.

Clearly, model-based sequential designs also fall under the category of conditionally optimal designs. In general terms, sequential designs aim at maximizing the expectation of a model-based criterion or utility function $g(\hat{y}(\mathbf{x};\theta, s^{p}))$ where $\hat{y}(\mathbf{x};\theta, s^{p})$ is the estimated value of the function of interest $y(\mathbf{x})$ under a statistical model with parameters $\theta$ and $s^p=\{(\bx_1,y(\bx_1)),\ldots,(\bx_p,y(\bx_p))\}$ is the set of observations. The $p^\text{th}$ design point is added by maximizing $E(g)$ conditional on the previous $p-1$ points, 
\begin{equation}
x_p = \text{argmax}\int_{\mathcal{X}}g(\hat{y}(\mathbf{x};\hat{\theta}_{p-1}))d\bx,
\end{equation}
where 
\begin{equation}
\hat{\theta}_{p-1}=\text{E}(\theta\mid s^{p-1}).
\end{equation}
is the model parameter estimate based on the $p-1$ observations. Examples of utility functions are the Shannon information in Bayesian D-optimal designs \citep{Lin56, Sto59, Ber79}, integrated mean squared error (IMSE) and maximum mean squared error (MMSE) \citep{SacSch88,SacSchWel89,SacWelMit89}, and other criteria specifically defined for optimization purposes \citep{JonSchWel98, ScoFraPow2011}. See also \cite{ChaVer95} for a review of Bayesian design criteria.

\begin{figure*}[t]
	\centering
	\begin{subfigure}[b]{0.48\textwidth}
		\centering
		\includegraphics[width=\textwidth]{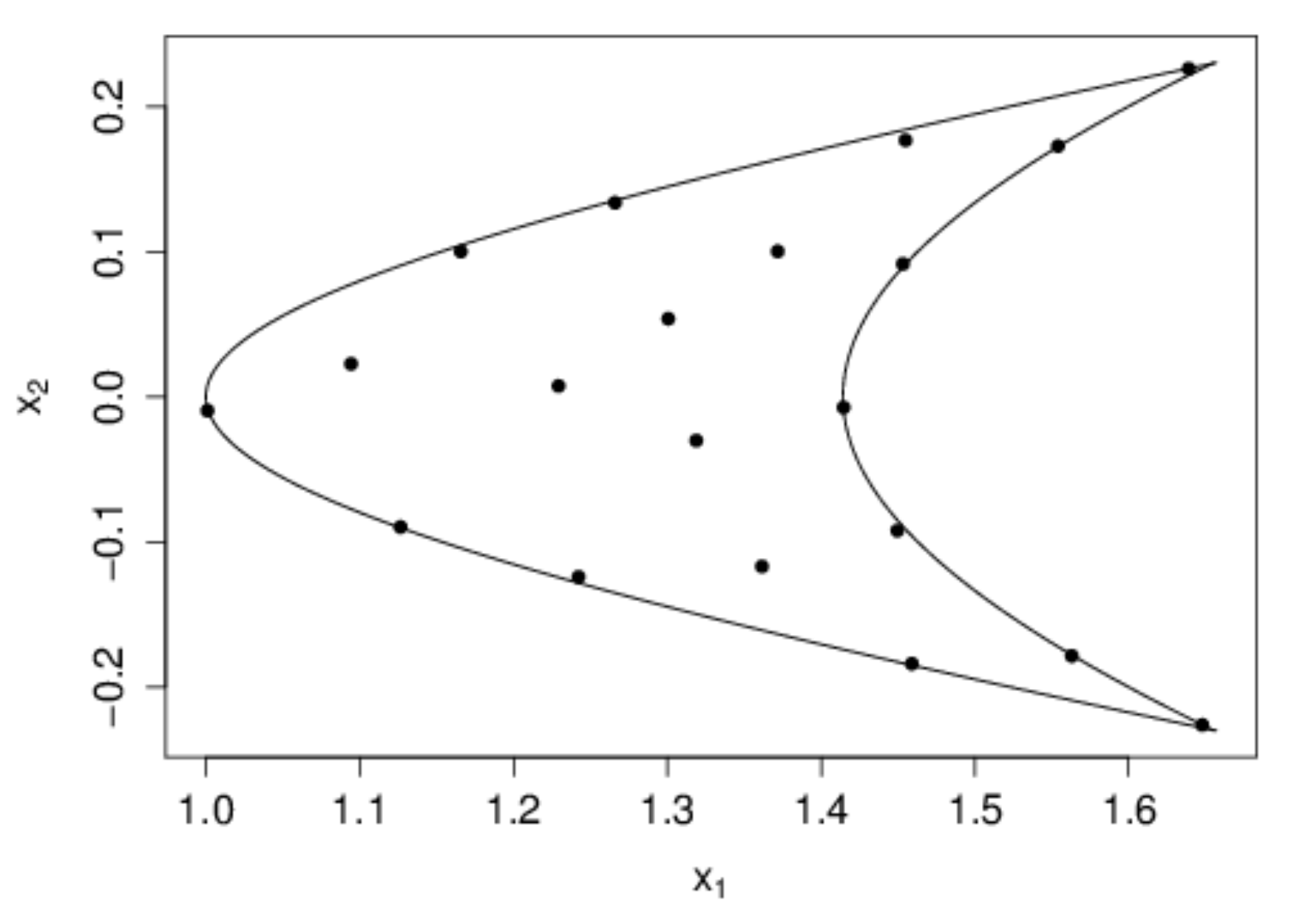}
		\caption{cMm design}
		\label{fig:crescent_cMm}
	\end{subfigure}
	\begin{subfigure}[b]{0.48\textwidth}
		\centering
		\includegraphics[width=\textwidth]{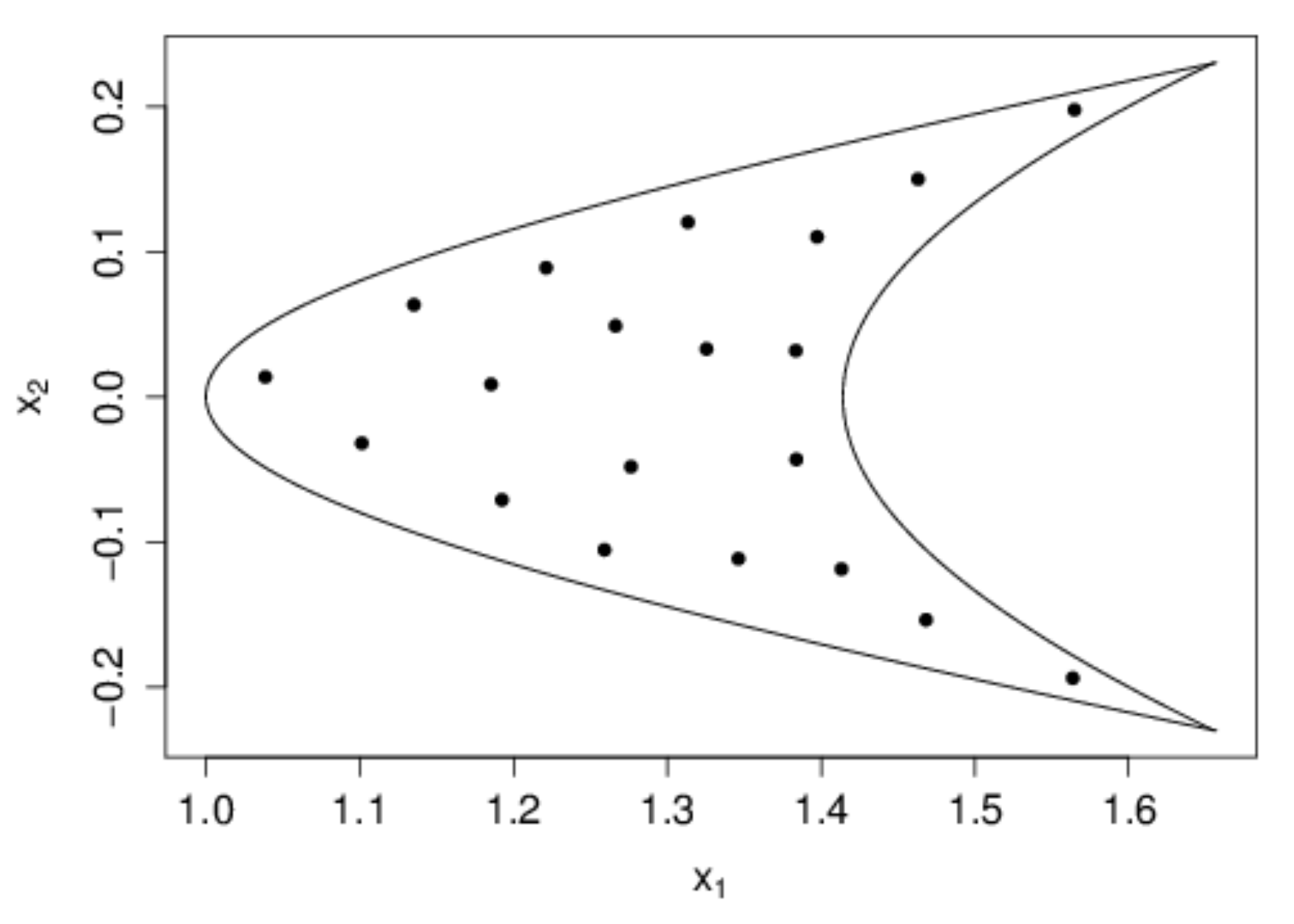}
		\caption{clustering design}
		\label{fig:crescent_fff}
	\end{subfigure}
	\caption{Space filling designs of size 20 on the 2d sub-space of $\mathcal{R}^2$ given in~(\ref{eqn:crescent}), obtained by (a) the cMm design algorithm (b) FFF design algorithm.}\label{fig:2d_ex}
\end{figure*}

\begin{figure*}[t]
	\centering
	\begin{subfigure}[b]{0.48\textwidth}
		\centering
		\includegraphics[width=\textwidth]{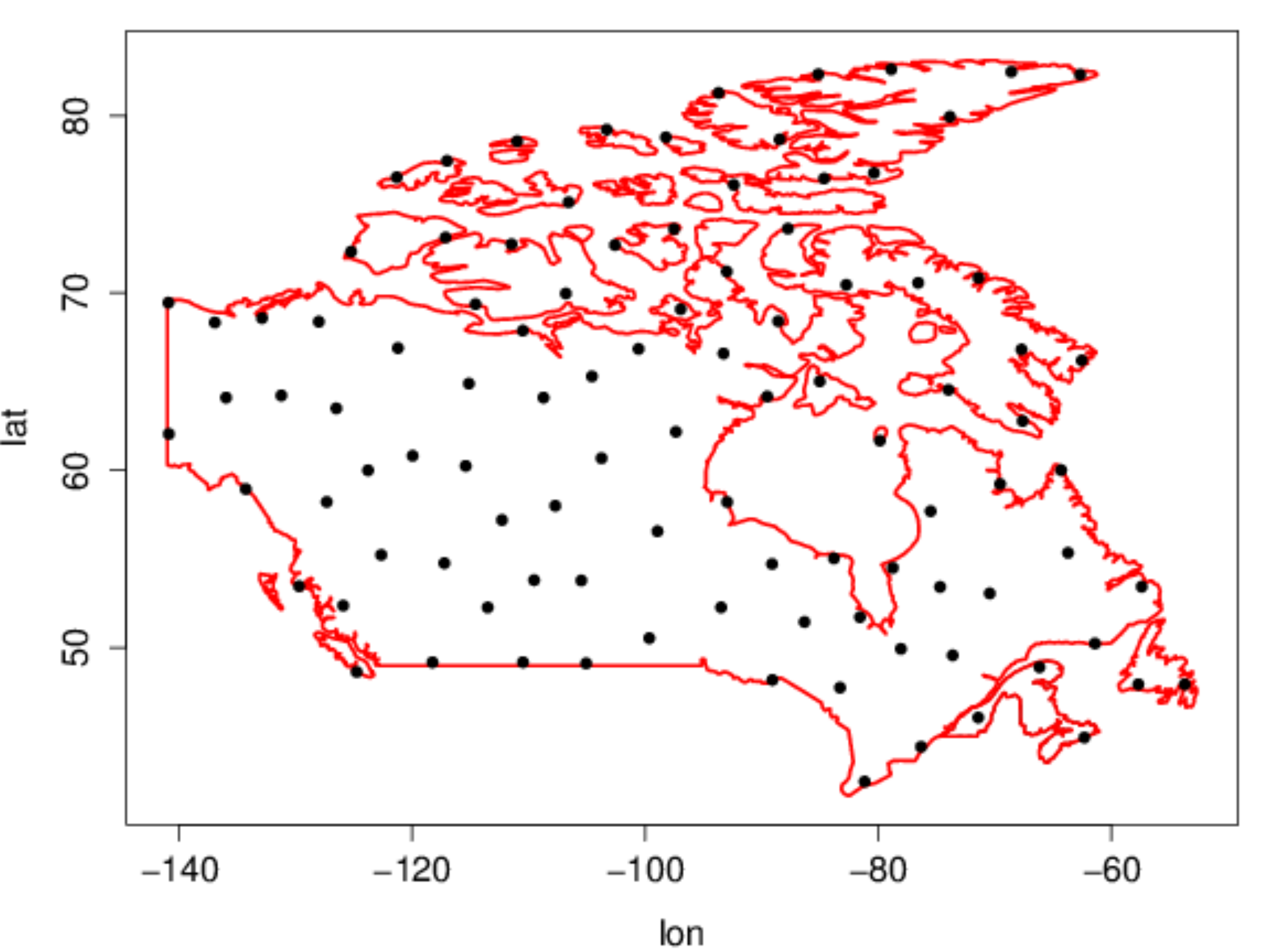}
		\caption{cMm design}
		\label{fig:Canada_cMm}
	\end{subfigure}
	\begin{subfigure}[b]{0.48\textwidth}
		\centering
		\includegraphics[width=\textwidth]{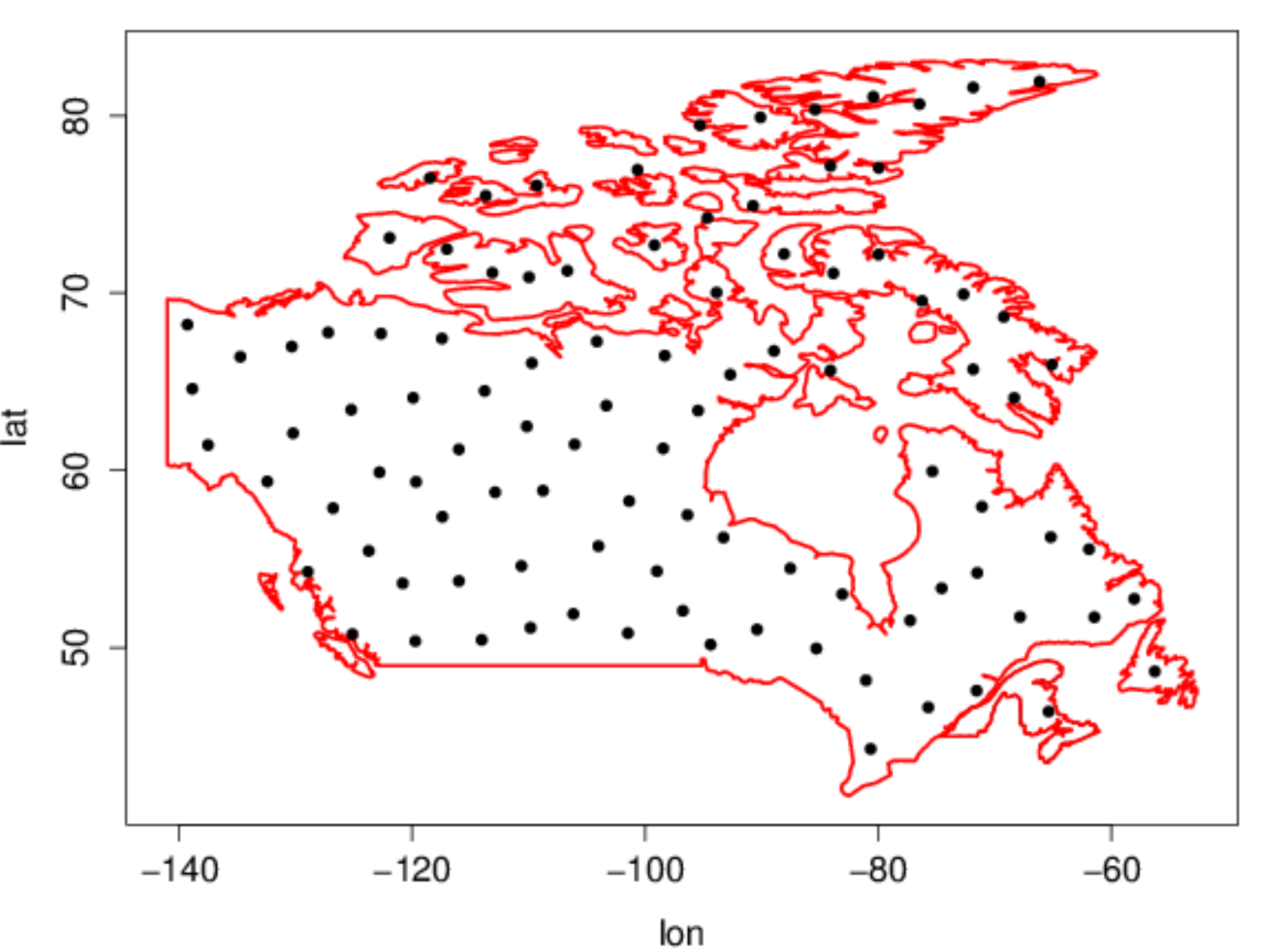}
		\caption{clustering design}
		\label{fig:Canada_fff}
	\end{subfigure}

	\caption{Space filling designs of size 100 over the map of Canada, obtained by (a) the cMm design algorithm (b) FFF design algorithm.}\label{fig:Canada_fff}
\end{figure*}

In our examples (Figures~\ref{fig:crescent_cMm} and \ref{fig:Canada_cMm}) we use the cMm design criterion with the following weighted Euclidean distance metric,
\begin{equation}
\label{wEuc}
\delta_{\boldsymbol{\omega}}(\bx_i,\bx_j)=\left(\sum_{d=1}^D \omega_d(x_{di}-x_{dj})^2\right)^{-\frac{1}{2}}
\end{equation}
where $\boldsymbol{\omega}=(\omega_1,\ldots,\omega_D)$ allows weighting the distance differently with respect to different dimensions or obtain the distance and the corresponding design in a subspace of ${\cal X}$ by setting  some $\omega_d$ equal to zero.  See \cite{JohMooYlv1990, LoeMooWil2010} for benefits of using this weighting in practice. 

\begin{algorithm}[H]
	\caption{Conditionally maximin design}\label{alg:design}
	\begin{algorithmic}[1]
	\renewcommand{\algorithmicrequire}{\textbf{Input:}}
	\renewcommand{\algorithmicensure}{\textbf{Return:}}
	\Require $S$\\
	
	Initialize the design: 
	\begin{itemize}
	\item[{\scriptsize1-1:}] sample $\bx^1$ from $S$;
	\item[{\scriptsize1-2:}] $s^1=\{\bx^1\}$;
	\item[{\scriptsize1-3:}] $\psi_i^1\gets \delta_{\theta}(\bx_i,\bx^1)$, for $\bx_i\in S$, $i=1,\ldots,N$.
	\end{itemize}
	\For{$p:=2, \ldots, P$} 
	\For{$i:=1, \ldots, N$} 
	\begin{itemize}
	\item[{\scriptsize3-1:}]  $\delta_i\gets \delta_{\theta}(\bx_i,\bx^{p-1})$, for $\bx_i\in S$, $i=1,\ldots,N$;
	\item[{\scriptsize3-2:}] $\psi_i^p\gets \min(\delta_i,\psi_i^{p-1})$,  $i=1,\ldots,N$;
	%\item[{\scriptsize3-3:}] $W_i^p\gets \exp{(-\frac{1}{\psi_i^p})}$,  $i=1,\ldots,N$;
	\end{itemize}
	\EndFor\\
	
	$\bx^p \gets \bx_{i_\text{max}}$ where $i_\text{max}$ is the index of the largest $\psi_i^p$.
	
	\EndFor
	
	\Ensure Design $s=\{\bx^1,\ldots,\bx^P\}$.
	\end{algorithmic}
	\end{algorithm}
	
\subsection{Alternative Design Algorithms}
	
Aside from the general family of conditionally optimal designs, having a uniformly covering sample of points over high-dimensional, highly constrained continuous input spaces makes possible the implementation of many existing design algorithms that would be otherwise infeasible or extremely expensive to use. Examples of these existing methods are algorithms for constructing D-optimal designs \citep{Mit1974, CooNac1980, Wel1984, NguMil1992, Mor2000} in classical design of experiments and the hierarchical clustering method of \cite{LekJon2014}.   

Since we will use the design algorithm of \cite{LekJon2014} to generate FFF designs for two of our examples we briefly review their method in the following.  \cite{LekJon2014} proposed to generate a Monte Carlo sample over the region of interest. As we discussed earlier this first step is not trivial for highly constrained regions. However, assuming that a Monte Carlo sample is already given the FFF designs are constructed by hierarchical clustering of the sample points where the number of clusters is equal to the number of design points. The design points are then determined as a summary of the clusters for example the centroids. Figure~\ref{fig:crescent_fff} shows an FFF design of size 20 generated on the crescent by taking the cluster centroids.

The choice of the summary or a point that represents the cluster in the design is important and depends on the target region. For example, \cite{LekJon2014} point out that choosing the cluster centroids could be specially inappropriate on non-convex regions since it could result in design points that fall outside the target space. We visualize this problem in Figure~\ref{fig:Canada_fff} where an FFF design of size 100 is generated as cluster centroids on the map of Canada. Combining two small islands into one cluster can result in design points that fall off the land. To resolve this issue \cite{LekJon2014} propose using a representative point from each cluster according to some distance based criterion such as the Maxpro criterion.

\subsection{Manifold designs}
We now revisit Example 3 where the input space is assumed to be  a manifold in $\mathcal{R}^3$. To obtain a space-filling design on a manifold we need to consider the fact that the target region lives in a space with lower dimensions than that of the sampling space and therefore is based on a different coordinate system. Optimizing a  distance-based design criterion with the Euclidean distance in $\mathcal{R}^3$ can result in a non-uniform design since the definition of closeness is different on the manifold than that given by the Euclidean distance in the space that it is embedded in. Note that the definition of the deviation function in terms of the Euclidean distances is not an issue in the sampling step since sampling is performed in the embedding Euclidean space.

Related work in the literature is that of \cite{PraHarBin16} who consider design and analysis of computer experiments on non-convex input regions. They use techniques in manifold learning to map the input space into a higher dimensional space where the Euclidean metric can be used. A summary of the method is as follows. 

Geodesic distances are obtained as the shortest paths between two points on a graph over the input region. The pairwise geodesic distances are then used in multidimensional scaling to find the representation of the sample points in a higher dimensional Euclidean space such that the geodesic distances are approximately preserved. The reason for this mapping is that merely replacing the Euclidean distance with the geodesic distance can result in non-positive definite covariance matrices in the Gaussian process model. 

\begin{figure*}
	\centering
	\includegraphics[width=.7\textwidth]{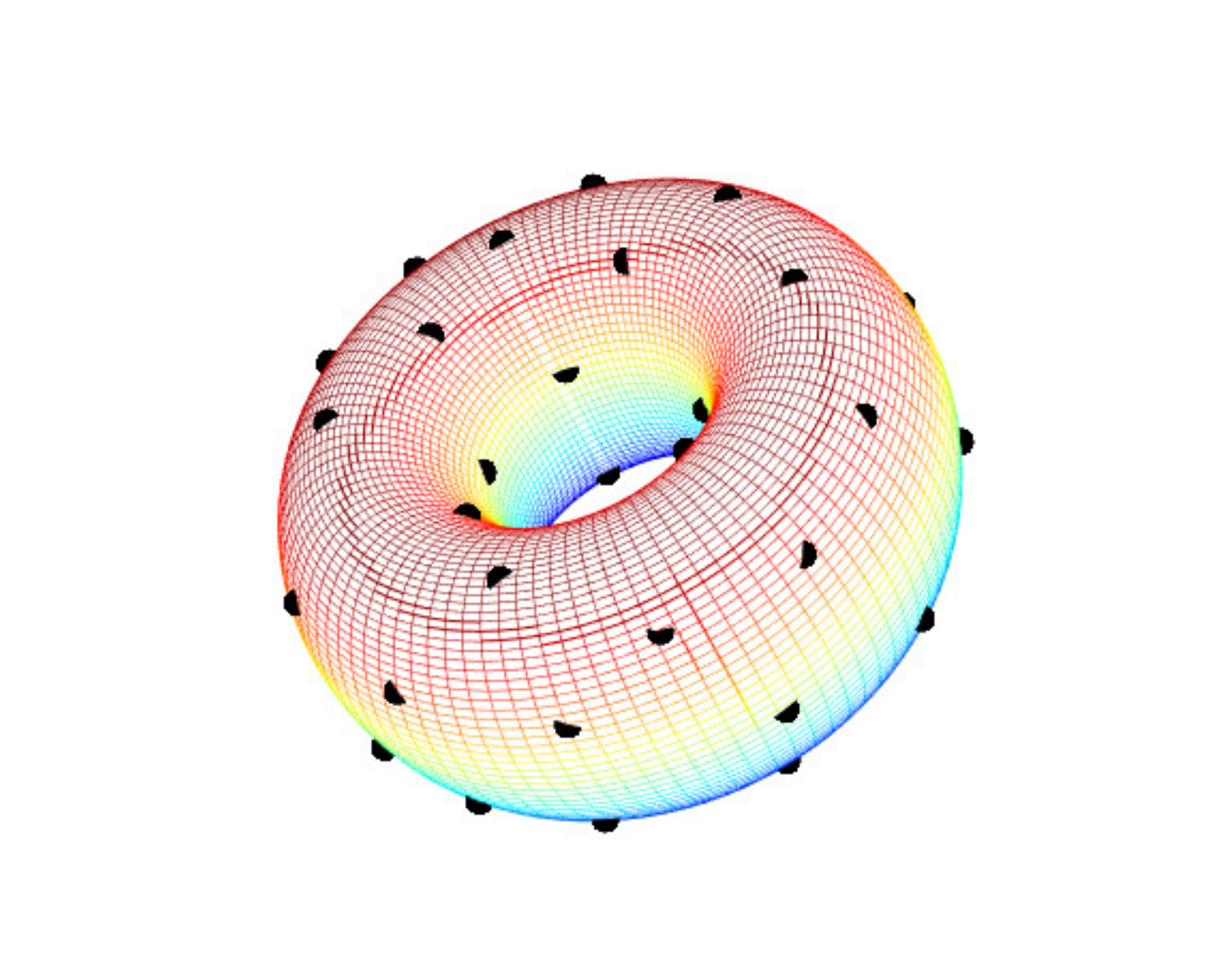}
		\caption{Conditionally maximin design on the surface of a torus. The distance metric used is the geodesic distance. }\label{fig:torus_geodesic}
	\end{figure*}

From a design point of view, however, using geodesic distances instead of the Euclidean distance would address the issue of appropriate distance metric. Figure~\ref{fig:torus_geodesic} shows a design of size 50 generated on the surface of a torus using the geodesic distances that are approximated using the Isomap algorithm based on the SCMC sample generated on the manifold. 

\section{Discussion}\label{sec:discussion}

In this paper we have proposed a sampling algorithm to generate a set of candidate points on any arbitrarily constrained region for Monte-Carlo based design algorithms. This sampling algorithm that takes advantage of the efficiency of the sequential Monte Carlo samplers is an exceptional tool for Monte Carlo sampling from constrained regions. A finite  uniform sample of points over high-dimensional, highly constrained continuous input spaces facilitates the implementation of existing design algorithms that would be otherwise difficult. Examples of these existing methods are algorithms for constructing D-optimal designs and the hierarchical clustering method of \cite{LekJon2014}. 

Starting with a simple example that is used to explain detailed implementation of the sampling algorithm, we demonstrate the performance of the proposed sampler by generating a sample over the map of Canada which is considered a highly constrained and difficult region from a sampling point of view. As a different challenging situation we also consider sampling over manifolds by generating points in the embedding higher-dimensional space.

To construct designs on these example regions we consider a general family of algorithms that are based on approximate optimization of a given design criterion by selecting points one at a time. Since the design criterion is optimized at each step given the design points that are selected up to the current step we refer to the design constructed by these type of algorithms as conditionally optimal designs. The computational gain that results from conditional optimization of distance based design criterion is what makes these family of algorithms preferable. We generate designs on our example surfaces by conditionally optimizing the maximin criterion as well as using the FFF designs proposed by \cite{LekJon2014}.

A challenging design scenario is creating a space-filling design on a manifold that arise in applications such as mixture experiments. The SCMC algorithm is an effective tool that generates samples that are within a controlled threshold from the manifold by sampling points in the embedding space. To generate a conditionally maximin design on the manifold, using the geodesic distance as the distance metric is recommended that is approximated as the shortest path on a graph constructed from the SCMC sample. 

Our contribution can be summarized into the following: Our adaptation of the SCMC sampling algorithm provides a discretization for high-dimensional, constrained, continuous input spaces that facilitates various existing design algorithms. In addition, we recommend a sequential selection algorithm that is adaptable to use various distance-based and model-based design criteria and is an efficient alternative to many existing methods.

\centerline{\bf ACKNOWLEDGEMENTS}

\renewcommand{\baselinestretch}{1.0} The research of Loeppky was supported by  Natural Sciences and Engineering Research Council of Canada Discovery Grant  ({RGPIN-2015-03895 }). The authors also acknowledge the support and encouragement of C. C. Essix.  

%\medskip

\baselineskip 0.7\normalbaselineskip
\bibliographystyle{asa}
\linespread{1}
\bibliography{psn}

\end{document}